\documentclass[preprintnumbers,nofootinbib,a4paper,superscriptaddress]{revtex4-2}

\usepackage{amsmath,latexsym,amssymb,amsfonts}
\usepackage{amsthm}
\usepackage{graphicx}
\usepackage{epstopdf}
\usepackage{graphics}
\usepackage{bm}
\usepackage{tikz}
\usepackage{tikz-cd}
\usetikzlibrary{tikzmark, arrows.meta}
\usepackage{enumitem}
\usepackage{mathrsfs}

\usepackage{xcolor}
\usepackage[normalem]{ulem}

\usepackage{cancel}
\usepackage[hidelinks]{hyperref}

\theoremstyle{definition}

\begin{document}

\title{\texorpdfstring{\textit{Twinor Theory}: Field Equations from $(A)dS_4^{\mathbb C}$ Spin-Tractors\\[0.4em]
\large A Twin to Twistor Theory Modelled on $(A)dS_4^{\mathbb C}$}{Twinor Theory: Field Equations from Complexified (A)dS Spin-Tractors}}

\author{Chen-Hsu Chien}
\email[E-mail: ]{chien@math.muni.cz}
\affiliation{Department of Mathematics and Statistics, Masaryk University}

\begin{abstract}
Twinor theory is formulated as a spacetime-based counterpart to twistor theory using holomorphic reductive Cartan geometry modelled on complexified $(A)dS_4$. The fundamental spin-tractor bundle defines Dirac twinors, its symmetric powers define ladder twinors, and their extremal Lorentz components define chiral twinors. The general \textit{twinor field equation} is $D\mathbf Y=\mathbf J$. Its adjacent contractions and mixed divergence are constrained by source relations; no additional equation is imposed on its completely symmetric projection. On a flat torsion-free Lorentzian background, the adjacent equations with the mass-normalization source and vanishing additional contracted sources, together with vanishing mixed divergence, reproduce the massive \textit{Dirac--Fierz--Pauli system}; the massless chiral realization gives the zero-rest-mass equations. Within the fixed mass normalization, the reality condition yielding a positive real mass squared is the \textit{anti-de Sitter reality condition} on the particle scale, independently of the cosmological scale and background curvature. The mass-coupling phase is therefore present in the classical connection before quantization. Nonlinear gravity is formulated through the full adjoint curvature, with energy--momentum and spin currents specifying Einstein--Cartan dynamics without requiring an action. The adjoint Bianchi identity constrains source compatibility, while a BGG-type sequence organizes torsion-free infinitesimal deformations. Matter interactions enter through the coupled connection and contracted source relations, with curvature compatibility recorded by $D\mathbf J=\mathcal F_D\mathbf Y$. Both chiralities are retained, and the additional consistency requirements for interacting higher-spin fields are distinguished from the underlying bundle construction.
\end{abstract}

\maketitle
\hypersetup{pdftitle={Twinor Theory: Field Equations from Complexified (A)dS Spin-Tractors},pdfauthor={Chen-Hsu Chien}}
\newpage
\section{Introduction}\label{sec:introduction}

Penrose's twistor programme was introduced as an approach to reconciling quantum theory and gravitation through a complex-geometric reformulation of spacetime and its fields \cite{Penrose1967,PenroseMacCallum1973}. In this description, spacetime points arise from incidence relations in twistor space, while null geometry gives conformal structure a fundamental role \cite{PenroseRindler1986}. The resulting framework encodes the conformal class without selecting a preferred metric scale. Its holomorphic structure supports the Penrose transform, which represents massless field solutions by suitable cohomology classes \cite{BastonEastwood1989}. Building on this framework, the nonlinear graviton and Penrose--Ward constructions relate holomorphic geometry to self-dual gravitational and gauge fields \cite{Penrose1976,Ward1977}. These correspondences motivate the search for a common geometric description of spacetime and its fields.

Two features suggest a complementary construction. Incorporating the opposite interacting chirality into the nonlinear twistor description of gravity is the historical \textit{googly problem} \cite{Penrose2015}; extensions include ambitwistor descriptions of non-self-dual gauge fields and twistor actions for full gauge and conformal gravitational theories \cite{Isenberg1978,Witten1978,Mason2005}. Particle masses and the cosmological constant also introduce scales not fixed by a conformal class alone. Massive particles admit multi-twistor descriptions \cite{Tod1977}, but a related question can be posed directly on spacetime: can massive and massless fields, together with nonlinear gravity, be organized by a common spinor geometry retaining both chiralities and independent particle and cosmological scales?

Cartan geometry and tractor calculus provide a natural setting for this question \cite{Sharpe1997,BaileyEastwoodGover1994,AttardFrancois2016_1,AttardFrancois2016_2}. Associated connections supply differential operators and curvature identities, while Bernstein--Gelfand--Gelfand sequences organize invariant operators and their compatibility on appropriate backgrounds \cite{CapSlovakSoucek2001}. The construction of local twistors from holomorphic conformal Cartan geometry is discussed in the companion work \cite{Chien2026}. Here the conformal Cartan model is replaced by $(A)dS_4^{\mathbb C}$, and a corresponding framework is termed \textbf{twinor theory}. The name expresses both its intended relationship to twistor theory and the paired chiral structure of its spin-tractor representations. No twistor-space transform or resolution of the googly problem is asserted.

The complexified homogeneous model admits the spin presentation
\begin{equation}
(A)dS_4^{\mathbb C}
\cong\mathrm{SO}(5,\mathbb C)/\mathrm{SO}(4,\mathbb C)
\cong\mathrm{Sp}(4,\mathbb C)\big/
\bigl(\mathrm{SL}(2,\mathbb C)_L\times\mathrm{SL}(2,\mathbb C)_R\bigr).
\end{equation}
For $G=\mathrm{Sp}(4,\mathbb C)$ and $H=\mathrm{SL}(2,\mathbb C)_L\times\mathrm{SL}(2,\mathbb C)_R$, the Cartan connection combines the Lorentz spin connections and soldering form with a cosmological scale $\ell$. The fundamental twinor (spin-tractor) bundle, its symmetric powers, and, for $s>0$, their extremal subbundles are
\begin{equation}
\mathcal T:=\mathcal P\times_H\mathbb C^4\cong S^A\oplus S_{A'},\qquad
\mathcal T_Y^{(2s)}:=\bigodot^{2s}\mathcal T,\qquad
\mathcal T_W^{(2s)}:=\bigodot^{2s}S^A\oplus\bigodot^{2s}S_{A'}.
\end{equation}
Sections of these bundles are Dirac, general ladder, and general chiral twinors, respectively. The ladder contains $2s+1$ chiral spinor summands, coupled by the soldering terms of its induced connection; for $s>0$, the chiral bundle retains the two extremal summands.

A particle scale $\lambda$ enters a scale-deformed twinor connection through the soldering coefficient $i/(2\sqrt2\lambda)$. The general \textit{twinor field equation} is $D\mathbf Y=\mathbf J$. Source relations constrain the adjacent contractions and mixed divergence; no additional field equation is imposed on the completely symmetric projection. On a flat torsion-free Lorentzian background, the adjacent equations with the mass-normalization source and vanishing additional contracted sources, together with the mixed-divergence constraints, reproduce the Dirac--Fierz--Pauli system \cite{Dirac1936,FierzPauli1939,BargmannWigner1948}. The massless chiral realization imposes the zero-rest-mass equations; selecting the extremal fields is additional to taking the massless limit of the full ladder. Internal representations are included by tensoring with an associated gauge bundle; for spin one, the chiral two-form encoding gives the Maxwell and Yang--Mills field strengths when the appropriate curvature and field equations are imposed.

In the fixed mass normalization, the massive equations give $m^2=\lambda^{-2}$. Positive real mass squared therefore selects $\lambda\in\mathbb R\setminus\{0\}$, the anti-de Sitter reality condition for the particle connection. The imaginary coefficient of the mass coupling is thus already present in the classical soldering term. The particle mass is determined by $\lambda$, whereas the cosmological constant is determined independently by $\ell$ through $\Lambda=-3/\ell^2$. Thus \textit{the particle reality condition does not fix the sign of $\Lambda$ or require a locally anti-de Sitter spacetime.} Nor does the classical coupling phase determine quantum commutation relations or a state space.

Nonlinear gravity is encoded by the full adjoint twinor curvature, including torsion. Its Bianchi identity is separated from the source relations specifying Einstein--Cartan dynamics: energy--momentum sources the Einstein three-form, and the Lorentz spin current sources torsion. These equations can be imposed directly without a variational derivation. The adjoint BGG-type sequence organizes torsion-free infinitesimal deformations \cite{Cap2026}; the linearized Einstein condition selects gravitational perturbations within this geometric description. Additionally, the MacDowell--Mansouri action and a Lorentz-valued auxiliary $BF$ formulation are retained as equivalent gravitational formulations \cite{MacDowellMansouri1977,FreidelStarodubtsev2005}. The deformation operators also organize classical gauge data relevant to quantization on curved backgrounds, which requires additional causal, algebraic, and state structures.

For coupled matter, $\mathbf J$ records the full first-order derivative data, and exterior differentiation gives $D\mathbf J=\mathcal F_D\mathbf Y$. This identity is available for every spin on a background of the stated Cartan type. It is automatic when $\mathbf J=D\mathbf Y$ is unconstrained; once source relations are imposed, it supplies compatibility conditions for the resulting equations. Retaining the symmetric derivative avoids imposing parallel transport, but does not by itself remove the curvature compatibility conditions of constrained higher-spin systems \cite{Buchdahl1958}.

Lorentzian real slices have signature $(+---)$. Complexified tangent vectors and spinor indices follow the Penrose conventions \cite{PenroseRindler1984}:
\begin{equation}
\begin{gathered}
v^{AA'}=\frac{1}{\sqrt2}\sigma_a^{AA'}v^a,\qquad
v_A=v^B\epsilon_{BA},\qquad v^A=\epsilon^{AB}v_B,\\
\epsilon_{12}=\epsilon^{12}=1,\qquad
\epsilon^{AC}\epsilon_{BC}=\delta^A_B,\qquad
\epsilon^{AB}\epsilon_{AB}=2,\qquad
v^2=v^{AA'}v_{AA'}.
\end{gathered}
\end{equation}
Primed indices obey the same conventions and remain independent until a reality condition is imposed. Symmetrization has unit weight. Units $c=\hbar=1$ are used. The holomorphic construction is local; on a Lorentzian real slice, the resulting bundles and operators are also used for smooth physical fields, without assuming that every smooth solution extends holomorphically. The curvature convention and $\Lambda=-3/\ell^2$ are specified in Section~\ref{sec:hrcg}.

Section~\ref{sec:hrcg} develops the holomorphic reductive Cartan geometry, its reality conditions, and the equivalent gravitational actions. Section~\ref{sec:twinor_bundle} constructs the twinor representations and contracted operators, followed by free fields, nonlinear gravity and its deformations, sourced matter equations, and their physical interpretation.

\section{Holomorphic Reductive Cartan Geometry on \texorpdfstring{$(A)dS_4^{\mathbb C}$}{Complexified (A)dS4} Spacetime}\label{sec:hrcg}

\paragraph*{\textbf{Reductive Cartan Geometry}}
A reductive Cartan geometry of type $(G,H)$ over a manifold $M$ consists of a principal $H$-bundle $\pi:\mathcal P\to M$, a Cartan connection $\vartheta\in\Omega^1(\mathcal P,\mathfrak g)$, and a chosen $\operatorname{Ad}(H)$-invariant decomposition $\mathfrak g=\mathfrak h\oplus\mathfrak m$. The Cartan connection satisfies
\begin{equation}
R_h^*\vartheta=\operatorname{Ad}_{h^{-1}}\vartheta,\qquad \vartheta(X^*)=X,\qquad \vartheta_p:T_p\mathcal P\xrightarrow{\sim}\mathfrak g,
\end{equation}
for $h\in H$, $X\in\mathfrak h$, and $p\in\mathcal P$, where $X^*$ is the fundamental vector field generated by $X$. In the complex analytic setting, all bundle data and pointwise isomorphisms are holomorphic. Relative to the chosen reductive decomposition,
\begin{equation}
\vartheta=\varpi+\varrho,
\end{equation}
where $\varpi$ is a principal $H$-connection and $\varrho$ is a horizontal, $H$-equivariant soldering form, inducing $TM\simeq\mathcal P\times_H\mathfrak m$. Local fields are obtained by pullback along a local section $\sigma:U\to\mathcal P$. Throughout this work, the construction is restricted to a trivializing open subset $U\subset M$, on which the local Cartan connection is $\mathcal A:=\sigma^*\vartheta\in\Omega^1(U,\mathfrak g)$.

\subsection{Holomorphic Reductive Cartan Geometry in Spin Representation}
\paragraph*{\textbf{Holomorphic Reductive Cartan Geometry on $(A)dS_4^{\mathbb C}$}}
A holomorphic Cartan geometry of type $(\mathrm{SO}(5,\mathbb C),\mathrm{SO}(4,\mathbb C))$ on a complex four-manifold is modelled on complexified $(A)dS_4$ (see \cite{Sharpe1997} for Cartan geometry).\footnote{Complexified Minkowski space $\Lambda=0$ is the homogeneous model of complex Poincar\'e Cartan geometry $(\mathrm{ISO}(4,\mathbb{C}), \mathrm{SO}(4,\mathbb{C}))$.} Assuming a spin lift, the isomorphism $\mathrm{Spin}(5,\mathbb C)\simeq\mathrm{Sp}(4,\mathbb C)$ gives the reductive decomposition
\begin{equation}
\mathfrak{sp}(4,\mathbb C)=\bigl(\mathfrak{sl}(2,\mathbb C)_L\oplus\mathfrak{sl}(2,\mathbb C)_R\bigr)\oplus(\mathbb C^2\otimes\mathbb C^2),
\end{equation}
where the transvection subspace transforms in the $(\tfrac12,\tfrac12)$ representation. The nondegenerate soldering form determines the holomorphic metric $g=\epsilon_{AB}\epsilon_{A'B'}e^{AA'}\otimes e^{BB'}$. Primed and unprimed spinors are independent over $\mathbb C$.

For a constant $\ell\in\mathbb C\setminus\{0\}$, the local Cartan connection $\mathcal A\in\Omega^1(U,\mathfrak g)$ and its curvature $\mathcal F\in\Omega^2(U,\mathfrak g)$, defined by $\mathcal F^\alpha{}_\beta:=d\mathcal A^\alpha{}_\beta+\mathcal A^\alpha{}_\gamma\wedge\mathcal A^\gamma{}_\beta$, take the block form
\begin{equation}\label{CartanAF}
\mathcal A^\alpha{}_\beta=\begin{pmatrix}
\omega_L{}^A{}_B&\frac{i}{\sqrt2\ell}e^{AB'}\\
\frac{i}{\sqrt2\ell}e_{A'B}&\omega_R{}_{A'}{}^{B'}
\end{pmatrix},\qquad
\mathcal F^\alpha{}_\beta=\begin{pmatrix}
R_L{}^A{}_B-\frac{1}{2\ell^2}e^{AC'}\wedge e_{C'B}&\frac{i}{\sqrt2\ell}\tau^{AB'}\\
\frac{i}{\sqrt2\ell}\tau_{A'B}&R_R{}_{A'}{}^{B'}-\frac{1}{2\ell^2}e_{A'C}\wedge e^{CB'}
\end{pmatrix},
\end{equation}
where the curvature components consist of the chiral spin curvatures $R_{L,R} := \mathrm{d}\omega_{L,R} + \omega_{L,R} \wedge \omega_{L,R}$ and the torsion 2-form $\tau^{AA'} := \mathrm{d}e^{AA'} + \omega_L{}^A{}_B \wedge e^{BA'} + \omega_R{}^{A'}{}_{B'} \wedge e^{AB'}$. Vanishing torsion determines the holomorphic Levi--Civita spin connection, whereas $\mathcal F=0$ characterizes local equivalence to the homogeneous Cartan model $\mathrm{(A)dS}_4^\mathbb{C}$.

\paragraph*{\textbf{Reality Condition and Cosmological Constant}}
A Lorentzian real spacetime geometry is selected by imposing reality conditions. The chiral spin connections are required to be complex conjugate to one another, $\overline{\omega_L{}^A{}_B}=\omega_R{}^{A'}{}_{B'}$, while the soldering 1-form satisfies the Hermiticity condition $\overline{e^{AB'}}=e^{BA'}$. The phase of $\ell$ selects the real form and the cosmological constant is $\Lambda=-3/\ell^2$.\footnote{On the homogeneous model, $\mathcal F=0$ gives the tangent action of the spin curvature,
$R_L{}^A{}_B\delta^{A'}{}_{B'}-\delta^A{}_BR_R{}_{B'}{}^{A'}=\ell^{-2}e^{AA'}\wedge e_{BB'}$. The curvature sign is fixed by $(\nabla_X\nabla_Y-\nabla_Y\nabla_X-\nabla_{[X,Y]})v=-R(X,Y)v$; for torsion-free coordinate derivatives this reads $[\nabla_c,\nabla_d]v^a=-R^a{}_{bcd}v^b$. Consequently,
$R^a{}_{bcd}=-\ell^{-2}(\delta^a_cg_{bd}-\delta^a_dg_{bc})$
and
$R_{ab}:=R^c{}_{acb}=-3\ell^{-2}g_{ab}=\Lambda g_{ab}$.}
\begin{enumerate}
    \item \textbf{de Sitter slice:} For $\ell\in i\mathbb R\setminus\{0\}$, the connection takes values in $\mathfrak{sp}(1,1)\cong\mathfrak{so}(4,1)$, and $\Lambda=-3/\ell^2>0$.
    \item \textbf{Anti-de Sitter slice:} For $\ell\in\mathbb R\setminus\{0\}$, the connection takes values in a complex-basis realization of $\mathfrak{sp}(4,\mathbb R)\cong\mathfrak{so}(3,2)$, and $\Lambda=-3/\ell^2<0$. \footnote{The matrices need not have real entries in the chiral spinor basis; they become real in a basis adapted to the invariant real structure.}
\end{enumerate}
These reality conditions select the real Cartan type; local equivalence to its homogeneous model additionally requires $\mathcal F=0$.

\subsection{The \texorpdfstring{$(A)dS_4^{\mathbb C}$}{Complexified (A)dS4} Action and its Chirality}

The action formulations integrate holomorphic four-forms over a real four-dimensional physical slice $\Gamma\subset M$. The Cartan curvature gives the MacDowell--Mansouri action and an equivalent Lorentz-valued auxiliary $BF$ formulation.

\paragraph*{\textbf{MacDowell--Mansouri Action}}
The nonchiral MacDowell--Mansouri action \cite{MacDowellMansouri1977} is constructed by inserting the grading matrix $\Upsilon=\operatorname{diag}(\mathbb I_2,-\mathbb I_2)$ into the curvature pairing, with the trace taken in the four-dimensional spin representation:
\begin{align}
S_{\mathrm{MM}}[\mathcal A]
&=-\frac{i\ell^2}{16\pi G}\int_\Gamma\operatorname{Tr}\left(\mathcal F\wedge\mathcal F\,\Upsilon\right)\nonumber\\
&=\frac{1}{16\pi G}\int_\Gamma\Biggl[\underbrace{i\left(\Sigma_{AB}\wedge R_L{}^{AB}-\bar\Sigma_{A'B'}\wedge R_R{}^{A'B'}\right)}_{\text{Palatini / Einstein--Cartan}}
+\frac{i}{4\ell^2}\underbrace{\left(\Sigma^{AB}\wedge\Sigma_{AB}-\bar\Sigma^{A'B'}\wedge\bar\Sigma_{A'B'}\right)}_{\text{Cosmological constant}}\Biggr]\nonumber\\
&\quad+\frac{i\ell^2}{16\pi G}\int_\Gamma\underbrace{\left(R_L{}^{AB}\wedge R_{L\,AB}-R_R{}^{A'B'}\wedge R_{R\,A'B'}\right)}_{\text{Gauss--Bonnet / Euler term}}.
\end{align}
Here $\Sigma^{AB}:=e^{(A}{}_{C'}\wedge e^{B)C'}$ and $\bar\Sigma^{A'B'}:=e_C{}^{(A'}\wedge e^{|C|B')}$ are the chiral two-form bases; the bar labels the independent primed sector and denotes complex conjugation only after imposing Lorentzian reality conditions. Both $R_L{}^{AB}$ and $R_R{}^{A'B'}$ are obtained by raising the second index of the curvature acting on upper spinors. The off-diagonal torsion terms cancel identically, so the decomposition holds off shell. The insertion of $\Upsilon$ is invariant under $H$, the structure group of the chosen Cartan reduction. The orientation $\mathrm{vol}_g=e^0\wedge e^1\wedge e^2\wedge e^3$ is related via
\begin{equation}
i\bigl(\Sigma_{AB}\wedge R_L{}^{AB}-\bar\Sigma_{A'B'}\wedge R_R{}^{A'B'}\bigr)
=R(g,\omega)\,\mathrm{vol}_g,\qquad
i\bigl(\Sigma^{AB}\wedge\Sigma_{AB}-\bar\Sigma^{A'B'}\wedge\bar\Sigma_{A'B'}\bigr)
=24\,\mathrm{vol}_g.
\end{equation}
Here $R(g,\omega)$ is the scalar curvature of the metric-compatible Lorentz connection in the adopted Riemann convention. The non-topological part is therefore $(16\pi G)^{-1}\int_\Gamma(R(g,\omega) -2\Lambda)\, \mathrm{vol}_g$, with $\Lambda=-3/\ell^2$. The Euler term has no bulk variation; with characteristic-class normalization, its integral over a closed physical slice gives the Euler characteristic.

\paragraph*{\textbf{First-Order BF Action}}
A Lorentz-valued auxiliary two-form is introduced as
\begin{equation}
\mathcal B=\begin{pmatrix}B_L{}^A{}_B&0\\0&B_R{}_{A'}{}^{B'}\end{pmatrix}\in\Omega^2(\Gamma,(\operatorname{ad}\mathcal P)|_\Gamma).
\end{equation}
The auxiliary-field action is
\begin{align}
S_{BF}[\mathcal B,\mathcal A]
&=i\int_\Gamma\left[\operatorname{Tr}(\mathcal B\wedge\mathcal F\Upsilon)+\frac{\rho}{2}\operatorname{Tr}(\mathcal B\wedge\mathcal B\Upsilon)\right]\nonumber\\
&=-i\int_\Gamma\Biggl[B_L{}^{AB}\wedge\mathcal F_{L\,AB}-B_R{}^{A'B'}\wedge\mathcal F_{R\,A'B'}
+\frac{\rho}{2}\left(B_L{}^{AB}\wedge B_{L\,AB}-B_R{}^{A'B'}\wedge B_{R\,A'B'}\right)\Biggr],
\end{align}
where $\rho=8\pi G/\ell^2$. Variation with respect to $\mathcal B$ gives the algebraic equation $\mathcal F_{\mathfrak h}+\rho\mathcal B=0$. Substitution of $\mathcal B=-\rho^{-1}\mathcal F_{\mathfrak h}$ recovers $S_{\mathrm{MM}}$, establishing an equivalent auxiliary-field formulation of the MacDowell--Mansouri theory. Related symmetry-breaking $BF$ formulations are discussed in \cite{FreidelStarodubtsev2005}; here the Lorentz-valued auxiliary field implements the chosen reduction directly.

On a Lorentzian real slice, the two chiral sectors are related by complex conjugation. The chiral Lorentz connection and two-forms $\Sigma^{AB}(e)$ connect this formulation to Pleba\'nski gravity, where independent two-forms are constrained to arise from a soldering form \cite{Plebanski1977}. The corresponding canonical connection formulations underlie loop quantum gravity \cite{Ashtekar1986,Barbero1995,Holst1996}; neither construction requires a particular sign of the cosmological constant. The metric is determined by $e$, while the Lorentz connection remains an independent variational field. In vacuum, variation with respect to the Lorentz connection imposes the torsion-free condition, uniquely determining the Levi--Civita spin connection $\omega_{L,R}=\mathring\omega_{L,R}(e)$. With minimally coupled Dirac matter, the connection equation is sourced by the intrinsic spin current, giving
\begin{equation}
\omega_{L,R}=\mathring\omega_{L,R}(e)+\kappa_{L,R},\qquad
\tau^{AA'}=\kappa_L{}^A{}_B\wedge e^{BA'}+\kappa_R{}^{A'}{}_{B'}\wedge e^{AB'}.
\end{equation}
In Einstein--Cartan theory, the contorsion $\kappa_{L,R}$ is determined algebraically by the matter spin density. For minimally coupled Dirac matter, substituting this solution into the total gravity--Dirac action gives a classically equivalent Levi--Civita formulation with a local axial-current four-fermion interaction proportional to $G(\bar\psi\gamma^a\gamma_5\psi)(\bar\psi\gamma_a\gamma_5\psi)$ \cite{HehlEtAl1976}. This substitution retains the effects of torsion through the contact interaction; it does not impose vanishing contorsion.

\section{Twinor Theory from \texorpdfstring{$(A)dS_4^{\mathbb C}$}{Complexified (A)dS4} Tractors}\label{sec:twinor_bundle}

The holomorphic Cartan geometry defines Dirac, adjoint, ladder, and chiral twinor bundles. The equation $D\mathbf Y=\mathbf J$ is the \emph{twinor field equation}; its projection constraints and source relations specify the dynamics, while the adjoint curvature encodes gravity and torsion. The construction is holomorphic on $U$, with independent primed and unprimed spinors. Reality conditions are introduced for the physical realizations below.

\paragraph*{\textbf{Local Twinor Bundle $\mathcal T$ and Dirac Twinors}}
The holomorphic Cartan geometry of Section~\ref{sec:hrcg} determines the local twinor bundle
\begin{equation}
\mathcal T:=\mathcal P\times_H\mathbb C^4\cong S^A\oplus S_{A'},
\end{equation}
where the splitting is induced by restriction of the fundamental $\mathrm{Sp}(4,\mathbb C)$ representation to $H=\mathrm{SL}(2,\mathbb C)_L\times\mathrm{SL}(2,\mathbb C)_R$. A section $Y_d=(u^A,v_{A'})^{\mathsf T}$ is called a \textbf{Dirac twinor}. The induced twinor (spin-tractor) connection is $\nabla^{\mathcal T}=d+\mathcal A$, with curvature $(\nabla^{\mathcal T})^2Y_d=\mathcal FY_d$. Here ``tractor'' refers to the associated bundle of a $G$-representation restricted to $H$, equipped with the connection induced by the Cartan connection; no conformal structure is required.

For a constant $\lambda\in\mathbb C\setminus\{0\}$, the scale-deformed twinor connection is
\begin{equation}\label{TC}
\nabla^{\mathcal T}_\lambda
\begin{pmatrix}u^A\\v_{A'}\end{pmatrix}
:=
d\begin{pmatrix}u^A\\v_{A'}\end{pmatrix}
+\begin{pmatrix}
\omega_L{}^A{}_B&\dfrac{i}{2\sqrt2\lambda}e^{AB'}\\
\dfrac{i}{2\sqrt2\lambda}e_{A'B}&\omega_R{}_{A'}{}^{B'}
\end{pmatrix}
\begin{pmatrix}u^B\\v_{B'}\end{pmatrix}.
\end{equation}
The geometric connection is recovered at $\lambda=\ell/2$. Otherwise, $\lambda$ is independent of $\ell$, with its physical interpretation determined by the real-slice field equations.

The coefficients along the soldering one-forms are zero-forms. Their contracted and contraction-free parts are\footnote{Write $e^{AA'}=e^{AA'}{}_\mu dx^\mu$ and denote its inverse by $e_{AA'}{}^\mu$, so that $e^{AA'}{}_\mu e_{BB'}{}^\mu=\delta^A{}_B\delta^{A'}{}_{B'}$. The frame derivative is $\nabla_{AA'}:=e_{AA'}{}^\mu\nabla_\mu$. The inverse coefficients are distinguished from the lowered one-form $e_{AA'}$; no additional frame field is introduced.}
\begin{align}\label{eq:ptdt}
\nabla_{BB'}u^A+\frac{i}{2\sqrt2\lambda}\delta_B{}^A v_{B'}
&=\frac12\delta_B{}^A\left(\nabla_{CB'}u^C+\frac{i}{\sqrt2\lambda}v_{B'}\right)
+\underbrace{\epsilon^{AC}\nabla_{(B|B'|}u_{C)}}_{U_{BB'}{}^A},\\
\nabla^{BB'}v_{A'}+\frac{i}{2\sqrt2\lambda}\delta_{A'}{}^{B'}u^B
&=\frac12\delta_{A'}{}^{B'}\left(\nabla^{BC'}v_{C'}+\frac{i}{\sqrt2\lambda}u^B\right)
+\underbrace{\epsilon_{C'A'}\nabla^{B(B'}v^{C')}}_{V^{BB'}{}_{A'}}.
\end{align}
The symbol $\nabla$ denotes the covariant derivative, acting through $\omega_L$ and $\omega_R$ on unprimed and primed indices, respectively.

\paragraph*{\textbf{Adjoint Twinor Bundle and its Adjoint BGG}}
The invariant symplectic form identifies the adjoint twinor bundle as
\begin{equation}\label{eq:adjoint_symmetric_twinor}
\mathcal AM:=\mathcal P\times_H\mathfrak{sp}(4,\mathbb C)
\cong\bigodot^2\mathcal T
\cong\mathfrak{so}(TM)\oplus TM
\cong\left(\bigodot^2S^A\oplus\bigodot^2S_{A'}\right)\oplus(S^A\otimes S_{A'}).
\end{equation}
This is an identification of associated bundles. The geometric adjoint connection is induced from $\nabla^{\mathcal T}$ and acts on an adjoint-valued $j$-form by
\begin{equation}\label{eq:adjoint_exterior_derivative}
d^{\nabla^{\mathcal A}}\Xi
=d\Xi+\mathcal A\wedge\Xi-(-1)^j\Xi\wedge\mathcal A,
\qquad
(d^{\nabla^{\mathcal A}})^2\Xi
=\mathcal F\wedge\Xi-\Xi\wedge\mathcal F.
\end{equation}
In particular, $\mathcal F\in\Omega^2(U,\mathcal AM)$ is an adjoint-twinor-valued curvature and satisfies the exact Bianchi identity
\begin{equation}\label{eq:adjoint_bianchi}
d^{\nabla^{\mathcal A}}\mathcal F
=d\mathcal F+\mathcal A\wedge\mathcal F-\mathcal F\wedge\mathcal A=0.
\end{equation}

The block expansion is
\begin{equation}\label{eq:adjoint_bianchi_blocks}
d^{\nabla^{\mathcal A}}\mathcal F=
\begin{pmatrix}
d^\omega R_L{}^A{}_B
&
\dfrac{i}{\sqrt2\ell}\left(
d^\omega\tau^{AB'}
-R_L{}^A{}_C\wedge e^{CB'}
+e^{AC'}\wedge R_R{}_{C'}{}^{B'}
\right)
\\[2mm]
\dfrac{i}{\sqrt2\ell}\left(
d^\omega\tau_{A'B}
-R_R{}_{A'}{}^{C'}\wedge e_{C'B}
+e_{A'C}\wedge R_L{}^C{}_B
\right)
&
d^\omega R_R{}_{A'}{}^{B'}
\end{pmatrix}=0.
\end{equation}
Here $d^\omega$ is the exterior covariant derivative. The diagonal soldering contributions cancel against the off-diagonal torsion contributions; hence these blocks are precisely the differential and torsion Bianchi identities. Neither torsion-freeness nor flatness is assumed, and the identities alone impose no gravitational dynamics.

On a fixed torsion-free background, infinitesimal deformations are described by the connection
\begin{equation}\label{eq:adjoint-deformation-connection}
\widetilde\nabla_X^{\mathcal A}\chi
=\nabla_X^{\mathcal A}\chi-\mathcal F(X,\eta),
\qquad
\eta=\operatorname{pr}_{TM}\chi.
\end{equation}
The holomorphic analogue of the BGG-type construction of \v{C}ap~\cite[Theorems 2.5 and 2.8]{Cap2026} organizes these deformations into the bundles
\begin{equation}
\mathcal H^0=TM,\qquad
\mathcal H^1\cong\bigodot^2T^*M,\qquad
\mathcal H^j=
\ker\!\left(
\operatorname{Alt}:\Lambda^jT^*M\otimes\mathfrak{so}(TM)
\longrightarrow\Lambda^{j+1}T^*M\otimes TM
\right),
\qquad 2\leq j\leq4,
\end{equation}
where $\operatorname{Alt}(\Phi)=\Phi\wedge e$; in particular, $\mathcal H^2$ is the algebraic Riemann curvature bundle. The resulting differential sequence is represented by
\begin{equation}\label{eq:full-BGG-adjoint}
\xi\xrightarrow{\quad D^0\quad}\psi
\xrightarrow{\quad D^1\quad}\mathcal R
\xrightarrow{\quad D^2\quad}\mathcal K
\xrightarrow{\quad D^3\quad}\mathcal Q,
\end{equation}
with the entries belonging to $\Gamma(\mathcal H^0),\ldots,\Gamma(\mathcal H^4)$, respectively. Here $\psi$ represents a metric deformation $\delta g=2\psi$, and $(D^0\xi)_{ab}=\mathring\nabla_{(a}\xi_{b)}$ is the deformation generated by an infinitesimal diffeomorphism. The derivative $\mathring\nabla$ is the background Levi--Civita connection, and $D^0\xi=0$ is the Killing equation. The operator $D^1\psi$ represents curvature deformation under the BGG splitting identification. The remaining operators are covariant exterior differentiation on $\mathcal H^2$ and $\mathcal H^3$. A general $\psi$ need not belong to $\operatorname{im}D^0$.

The splitting operators $L^j:\Gamma(\mathcal H^j)\to\Omega^j(U,\mathcal AM)$ relate this sequence to adjoint-valued differentiation through
\begin{equation}
d^{\widetilde\nabla^{\mathcal A}}L^j=L^{j+1}D^j,
\qquad 0\leq j\leq3.
\end{equation}
The sequence is a complex on constant-sectional-curvature backgrounds, including the Cartan-flat model, but need not be a complex on a general torsion-free background. It organizes infinitesimal geometry; gravitational dynamics is imposed separately through the linearized Einstein equation.

\paragraph*{\textbf{General Ladder Twinors}}
For $2s\in\mathbb N_0$, the symmetric twinor bundle is
\begin{equation}
\mathcal T_Y^{(2s)}:=\bigodot^{2s}\mathcal T
\cong\bigoplus_{k=0}^{2s}\mathcal T^{(2s-k,k)}
\cong\bigoplus_{k=0}^{2s}
\left(\bigodot^{2s-k}S^A\otimes\bigodot^kS_{A'}\right).
\end{equation}
A \textbf{general ladder twinor} is a section $\mathbf Y\in\Gamma(U,\mathcal T_Y^{(2s)})$, represented for $2s>0$ by the chiral spinor multiplet
\begin{equation}
\mathbf Y=
\begin{pmatrix}
Y^{(2s,0)\,A_1\cdots A_{2s}}\\[1mm]
Y^{(2s-1,1)}{}_{A'_1}{}^{A_2\cdots A_{2s}}\\
\vdots\\
Y^{(0,2s)}{}_{A'_1\cdots A'_{2s}}
\end{pmatrix},
\qquad
\mathcal T^{(2s-k,k)}
\longleftrightarrow
\left(\frac{2s-k}{2},\frac{k}{2}\right).
\end{equation}
For $s=0$, the multiplet consists of the scalar component $Y^{(0,0)}$.

The induced connection acts on each symmetric twinor slot and couples adjacent summands through the soldering form. Its irreducible components are described by the decomposition of an arbitrary one-form $\Xi\in\Omega^1(U,\mathcal T_Y^{(2s)})$:
\begin{equation}\label{eq:source_irreducible_decomposition}
\begin{aligned}
\Xi^{(2s-k,k)}{}_{BB'A'_1\cdots A'_k}{}^{A_{k+1}\cdots A_{2s}}
={}&\epsilon_{CB}
(\operatorname{sym}\Xi^{(2s-k,k)})_{B'A'_1\cdots A'_k}{}^{CA_{k+1}\cdots A_{2s}}\\
&+\frac{2s-k}{2s-k+1}\delta_B{}^{(A_{k+1}}
(\operatorname{ctr}_L\Xi^{(2s-k,k)})_{B'A'_1\cdots A'_k}{}^{A_{k+2}\cdots A_{2s})}\\
&+\frac{k}{k+1}\epsilon_{BC}\epsilon_{B'(A'_1}
(\operatorname{ctr}_R\Xi^{(2s-k,k)})_{A'_2\cdots A'_k)}{}^{CA_{k+1}\cdots A_{2s}}\\
&-\frac{k(2s-k)}{(k+1)(2s-k+1)}
\delta_B{}^{(A_{k+1}}\epsilon_{B'(A'_1}
(\operatorname{ctr}_0\Xi^{(2s-k,k)})_{A'_2\cdots A'_k)}{}^{A_{k+2}\cdots A_{2s})}.
\end{aligned}
\end{equation}
Here $BB'$ labels the coefficient along $e^{BB'}$, and parentheses denote normalized symmetrization separately in each spinor type. The adjacent contractions, mixed contraction, and completely symmetric projection are
\begin{equation}\label{eq:four_twinor_components}
\begin{aligned}
(\operatorname{ctr}_L\Xi^{(2s-k,k)})_{B'A'_1\cdots A'_k}{}^{A_{k+2}\cdots A_{2s}}
&:=\Xi_{B(B'A'_1\cdots A'_k)}{}^{BA_{k+2}\cdots A_{2s}},\\
(\operatorname{ctr}_R\Xi^{(2s-k,k)})_{A'_2\cdots A'_k}{}^{CA_{k+1}\cdots A_{2s}}
&:=\epsilon^{(C|B|}\epsilon^{D'B'}
\Xi_{BB'D'A'_2\cdots A'_k}{}^{A_{k+1}\cdots A_{2s})},\\
(\operatorname{ctr}_0\Xi^{(2s-k,k)})_{A'_2\cdots A'_k}{}^{A_{k+2}\cdots A_{2s}}
&:=\epsilon^{D'B'}
\Xi_{BB'D'A'_2\cdots A'_k}{}^{BA_{k+2}\cdots A_{2s}},\\
(\operatorname{sym}\Xi^{(2s-k,k)})_{B'A'_1\cdots A'_k}{}^{CA_{k+1}\cdots A_{2s}}
&:=\epsilon^{(C|B|}
\Xi_{B(B'A'_1\cdots A'_k)}{}^{A_{k+1}\cdots A_{2s})}.
\end{aligned}
\end{equation}
The component superscript is suppressed on the right. These maps extract zero-form spinor coefficients, and terms requiring absent indices are omitted throughout. The notation $\operatorname{ctr}=(\operatorname{ctr}_L,\operatorname{ctr}_R)$ collects the adjacent contractions; $\operatorname{ctr}_0$ is stated separately.

Applying this decomposition to $\Xi^{(2s-k,k)}=\nabla Y^{(2s-k,k)}$ and including the soldering contributions from the Leibniz rule gives
\begin{equation}\label{eq:ladder_frame_derivative}
\begin{aligned}
&\left(\nabla^{\mathcal T_Y^{(2s)}}_{\lambda,BB'}\mathbf Y\right){}^{(2s-k,k)}{}_{A'_1\cdots A'_k}{}^{A_{k+1}\cdots A_{2s}}\\
&\quad=\epsilon_{CB}
\bigl(\operatorname{sym}(\nabla Y^{(2s-k,k)})\bigr)_{B'A'_1\cdots A'_k}{}^{CA_{k+1}\cdots A_{2s}}\\
&\qquad+\frac{2s-k}{2s-k+1}\delta_B{}^{(A_{k+1}}
\left[
\operatorname{ctr}_L(\nabla Y^{(2s-k,k)})
+\frac{i(2s-k+1)}{2\sqrt2\lambda}Y^{(2s-k-1,k+1)}
\right]{}_{B'A'_1\cdots A'_k}{}^{A_{k+2}\cdots A_{2s})}\\
&\qquad+\frac{k}{k+1}\epsilon_{BC}\epsilon_{B'(A'_1}
\left[
\operatorname{ctr}_R(\nabla Y^{(2s-k,k)})
+\frac{i(k+1)}{2\sqrt2\lambda}Y^{(2s-k+1,k-1)}
\right]{}_{A'_2\cdots A'_k)}{}^{CA_{k+1}\cdots A_{2s}}\\
&\qquad-\frac{k(2s-k)}{(k+1)(2s-k+1)}
\delta_B{}^{(A_{k+1}}\epsilon_{B'(A'_1}
\bigl(\operatorname{ctr}_0(\nabla Y^{(2s-k,k)})\bigr)_{A'_2\cdots A'_k)}{}^{A_{k+2}\cdots A_{2s})}.
\end{aligned}
\end{equation}
The bracketed expressions are the adjacent contractions of the full induced derivative. The soldering terms leave its mixed and completely symmetric projections unchanged. The subscript $k$ on a projection refers to the $(2s-k,k)$ component.

The induced connection already closes on the finite complex ladder multiplet before any reality condition is imposed. Reality conditions select its physical realization.

\paragraph*{\textbf{General Chiral Twinors}}
For $s>0$, the extremal components define the chiral bundle
\begin{equation}
\mathcal T_W^{(2s)}:=\bigodot^{2s}S^A\oplus\bigodot^{2s}S_{A'}
\subseteq\mathcal T_Y^{(2s)},\qquad
\mathbf W:=
\begin{pmatrix}
W^{(2s,0)\,A_1\cdots A_{2s}}\\
W^{(0,2s)}{}_{A'_1\cdots A'_{2s}}
\end{pmatrix}
\in\Gamma(U,\mathcal T_W^{(2s)}).
\end{equation}
A section $\mathbf W$ is called a \textbf{general chiral twinor}; its two entries have opposite chirality. The chiral bundle is an $H$-associated subbundle of $\mathcal T_Y^{(2s)}$, equipped with the spin connection induced by $\omega_L\oplus\omega_R$. When extracted from $\mathbf Y$, its entries are $W^{(2s,0)}=Y^{(2s,0)}$ and $W^{(0,2s)}=Y^{(0,2s)}$. The massless realization imposes the chiral field equations specified below; the bundle definition alone imposes no mass condition.

\paragraph*{\textbf{Field Strengths and Gauge Potentials}}
Let $Q\to U$ be a principal gauge bundle with complex structure group $K$ and Lie algebra $\mathfrak k$, and let $V$ be an associated vector bundle. For $s\geq1$, the soldering form defines the algebraic encoding
\begin{equation}\label{eq:chiral_twinor_curvature}
\begin{gathered}
\mathbf W\in\Gamma(U,\mathcal T_W^{(2s)}\otimes V),\qquad
F_{\mathrm{ch}}:=(\iota_s\otimes\mathbf1_V)\mathbf W
\in\Omega^2(U,\mathcal T_Y^{(2s-2)}\otimes V),\\
\begin{aligned}
F_L{}^{A_3\cdots A_{2s}}
&:=\frac12W^{(2s,0)\,AB A_3\cdots A_{2s}}
\epsilon_{A'B'}e_A{}^{A'}\wedge e_B{}^{B'},\\
F_R{}_{A'_3\cdots A'_{2s}}
&:=\frac12W^{(0,2s)}{}_{A'B'A'_3\cdots A'_{2s}}
\epsilon_{AB}e^{AA'}\wedge e^{BB'}.
\end{aligned}
\end{gathered}
\end{equation}
Internal indices are suppressed. These formulas define the embedding $\iota_s:\mathcal T_W^{(2s)}\hookrightarrow\Lambda^2T^*U\otimes\mathcal T_Y^{(2s-2)}$, with $\mathcal T_Y^{(0)}:=U\times\mathbb C$. For $s=1$, $F_{\mathrm{ch}}=F_L+F_R$ is a $V$-valued two-form; for $s>1$, the two contributions occupy distinct extremal coefficient summands, so $F_{\mathrm{ch}}=(F_L,0,\ldots,0,F_R)$.

For Yang--Mills fields, $s=1$ and $V=\operatorname{ad}Q:=Q\times_{\operatorname{Ad}}\mathfrak k$. A local connection potential $A\in\Omega^1(U,\mathfrak k)$ realizes the encoded field strength by the Yang--Mills curvature relation \cite{YangMills1954}
\begin{equation}\label{eq:twinor_potential_realization}
F:=F_{\mathrm{ch}}=dA+A\wedge A,\qquad
D^{\mathrm{ad}}F=dF+A\wedge F-F\wedge A=0.
\end{equation}
The second relation is the Bianchi identity, and both relations hold on curved backgrounds. Here $A\wedge A$ means $\tfrac12[A\wedge A]$ intrinsically, or matrix multiplication combined with the exterior product in a matrix representation. A one-dimensional Abelian gauge group gives the complexified electromagnetic case, with trivial adjoint bundle and $F=dA$.

For $s=2$ and trivial internal coefficients, $\mathcal T_W^{(4)}$ is identified with the Weyl subbundle of $\mathcal H^2$, encoding the Weyl component of either the torsion-free curvature or its BGG deformation representative. The remaining curvature components are not included. The geometric adjoint bundle $\mathcal AM$ is distinct from $\operatorname{ad}Q$.

The same algebraic encoding applies to the extremal components of a massive ladder twinor. A potential realization requires additional differential relations: a Proca field has $F=dA$, whereas higher-spin formulations require compatible field content and curvature constraints.\footnote{For massive higher-spin potential and field-strength formulations on $(A)dS_4$, see \cite{KhabarovZinoviev2019}.} The two-form encoding alone does not determine a gauge algebra or a potential.

\paragraph*{\textbf{Coupled Twinor Connection}}
For a representation $\rho_V:K\to\mathrm{GL}(V_0)$ of the internal gauge group, let $V:=Q\times_{\rho_V}V_0$ be the associated vector bundle, with induced connection $a_V=\rho_{V*}(A)$ and curvature $F_V=\rho_{V*}(F)=da_V+a_V\wedge a_V$. On $\mathcal E_s:=\mathcal T_Y^{(2s)}\otimes V$, the coupled connection and curvature are
\begin{equation}\label{eq:coupled_twinor_connection}
D=d+\mathcal A_\lambda^{(2s)}\otimes\mathbf1
+\mathbf1\otimes a_V,\qquad
\mathcal F_D=\mathcal F_\lambda^{(2s)}\otimes\mathbf1
+\mathbf1\otimes F_V.
\end{equation}
The superscript $(2s)$ denotes the induced representation. The gauge normalization, including coupling constants, is contained in $a_V$. Mixed curvature terms cancel because the two actions operate on separate tensor factors. For chiral fields, $\mathcal T_W^{(2s)}\otimes V$ is equipped with the Lorentz and internal gauge connections. The fundamental twinor curvature is
\begin{equation}\label{eq:particle_twinor_curvature}
\mathcal F_\lambda=
\begin{pmatrix}
R_L{}^A{}_B-\dfrac{1}{8\lambda^2}e^{AC'}\wedge e_{C'B}
&
\dfrac{i}{2\sqrt2\lambda}\tau^{AB'}
\\[2mm]
\dfrac{i}{2\sqrt2\lambda}\tau_{A'B}
&
R_R{}_{A'}{}^{B'}-\dfrac{1}{8\lambda^2}e_{A'C}\wedge e^{CB'}
\end{pmatrix}.
\end{equation}
Here $R_{L,R}$ and $\tau$ are the spin curvature and torsion. For several internal gauge factors, the same construction gives $\mathcal E_s=\mathcal T_Y^{(2s)}\otimes V_1\otimes\cdots\otimes V_N$: the connection and curvature contributions are summed over the factors, each acting as the identity on all other factors.

\paragraph*{\textbf{Exterior Degree and Covariant Differentiation}}
Twinor fields are bundle-valued zero-forms, independently of their spinor rank. The coupled derivative extends to bundle-valued forms and satisfies
\begin{equation}
D:\Omega^j(U,\mathcal E_s)\longrightarrow\Omega^{j+1}(U,\mathcal E_s),
\qquad
D^2\Theta=\mathcal F_D\wedge\Theta.
\end{equation}
The \emph{general twinor field equation} and its curvature compatibility identity are
\begin{equation}\label{eq:master_twinor_equation}
D\mathbf Y=\mathbf J,\qquad
D\mathbf J=\mathcal F_D\mathbf Y.
\end{equation}
Here $\mathbf J=e^{BB'}\mathbf J_{BB'}\in\Omega^1(U,\mathcal E_s)$ is the total source one-form, whose zero-form coefficients satisfy $\mathbf J_{BB'}=D_{BB'}\mathbf Y$ on solutions. The coefficients are given by \eqref{eq:ladder_frame_derivative} with $\nabla$ replaced by $\nabla+a_V$.

For the massive realization, the induced adjacent couplings are adjusted through the algebraic \emph{mass-normalization source}
\begin{equation}\label{eq:mass_correction}
\begin{aligned}
(\mathbf j_{\mathrm{mass}}[\mathbf Y])_{L,k}
&:=\frac{i(2s-k-1)}{2\sqrt2\lambda}Y^{(2s-k-1,k+1)},
&&0\leq k<2s,\\
(\mathbf j_{\mathrm{mass}}[\mathbf Y])_{R,k}
&:=\frac{i(k-1)}{2\sqrt2\lambda}Y^{(2s-k+1,k-1)},
&&1\leq k\leq2s.
\end{aligned}
\end{equation}
The decomposition \eqref{eq:source_irreducible_decomposition} uniquely lifts these coefficients to a one-form $\mathbf J_{\mathrm{mass}}[\mathbf Y]$ with
\begin{equation}\label{eq:mass_normalization_source}
\operatorname{ctr}\mathbf J_{\mathrm{mass}}=\mathbf j_{\mathrm{mass}},
\qquad
\operatorname{ctr}_0\mathbf J_{\mathrm{mass}}=0,
\qquad
\operatorname{sym}\mathbf J_{\mathrm{mass}}=0.
\end{equation}
Placing this source on the right-hand side of the twinor field equation leaves the effective coefficient $i/(\sqrt2\lambda)$ in both adjacent equations. It is linear and algebraic in $\mathbf Y$, introduces no independent field, and vanishes for $2s=1$; for $s=0$ it is absent. This source relation fixes the adjacent mass normalization while leaving the ladder bundle and its induced connection unchanged.

The source relations are
\begin{equation}\label{eq:total_twinor_source}
\operatorname{ctr}\mathbf J
=\mathbf j_{\mathrm{mass}}[\mathbf Y]+\mathbf j_{\mathrm{phys}},
\qquad
\operatorname{ctr}_0\mathbf J=\mathbf j_0,
\qquad
\operatorname{sym}\mathbf J=\operatorname{sym}(D\mathbf Y).
\end{equation}
The additional contracted sources are $\mathbf j_{\mathrm{phys}}$ and $\mathbf j_0$. The last equality records the symmetric derivative without imposing an additional equation; this projection remains subject to differential consequences of the coupled system.

Writing $\mathcal D:=\nabla+a_V$, subtraction of the mass-normalization source from the induced adjacent couplings gives
\begin{equation}\label{eq:sourced_adjacent_ladder}
\begin{aligned}
\operatorname{ctr}_L\bigl(\mathcal D Y^{(2s-k,k)}\bigr)
+\frac{i}{\sqrt2\lambda}Y^{(2s-k-1,k+1)}
&=j_{L,k},\\
\operatorname{ctr}_R\bigl(\mathcal D Y^{(2s-k-1,k+1)}\bigr)
+\frac{i}{\sqrt2\lambda}Y^{(2s-k,k)}
&=j_{R,k+1},
\end{aligned}
\qquad 0\leq k<2s.
\end{equation}
Here $j_{L,k}$ and $j_{R,k+1}$ are components of $\mathbf j_{\mathrm{phys}}$. The full source is reconstructed from \eqref{eq:source_irreducible_decomposition} and \eqref{eq:total_twinor_source}.

For massless chiral fields, $\mathbf Y$ is replaced by $\mathbf W$, and $\mathcal D$ is used without the particle soldering coupling or the mass-normalization source. The covariant sourced versions of \eqref{eq:ZRM} are obtained, subject to their curvature compatibility conditions.

\subsection{Free Fields and Gauge-Coupled Realizations}\label{sec:freefield_spin}

The holomorphic construction is restricted to a Lorentzian real slice. With the mass normalization in \eqref{eq:mass_correction}, the massive equations give $m^2=\lambda^{-2}$. Positive real mass squared selects the \textit{AdS reality condition} $\lambda\in\mathbb R\setminus\{0\}$ for the particle connection. The positive-mass branch $m=\lambda^{-1}>0$ is chosen below. For the free-field analysis, a fixed flat, torsion-free metric background is used, with matter backreaction neglected. In a parallel spin frame and compatible affine coordinates, $\omega_L=\omega_R=0$, $e$ is constant, and $\nabla_{AA'}=\partial_{AA'}$. Metric flatness is distinguished from Cartan flatness: the soldering contributions remain in the geometric and particle curvatures at finite $\ell$ and $\lambda$, respectively. For $\Lambda\neq0$, this background is therefore used only for the test-field analysis.

\paragraph*{\textbf{Dirac Fields from Dirac Twinors}}
For $2s=1$, the mass-normalization source vanishes. On the positive-mass branch, the vanishing adjacent contractions in \eqref{eq:ptdt} therefore give
\begin{equation}
\partial_{AA'}u^A+\frac{im}{\sqrt2}v_{A'}=0,\qquad
\partial^{AA'}v_{A'}+\frac{im}{\sqrt2}u^A=0.
\end{equation}
These are equivalent to the Dirac equation $(i\gamma^a\partial_a-m)Y_d=0$ in the compatible chiral representation, with $\{\gamma^a,\gamma^b\}=2g^{ab}\mathbf1$. The contraction-free components $U,V$ are not set to zero; imposing their vanishing additionally gives the stronger parallel-section equation $\nabla^{\mathcal T}_\lambda Y_d=0$.

\paragraph*{\textbf{Massive Fields from General Ladder Twinors}}
For $2s\in\mathbb N_{>0}$, the free massive spin-$s$ realization on $\mathbf Y\in\Gamma(U,\mathcal T_Y^{(2s)})$ has vanishing additional contracted sources, $\mathbf j_{\mathrm{phys}}=\mathbf j_0=0$, while retaining the mass-normalization source. Its adjacent projections satisfy
\begin{equation}\label{eq:DFP_full}
\operatorname{ctr}\!\left(\nabla^{\mathcal T_Y^{(2s)}}_\lambda\mathbf Y\right)
=\mathbf j_{\mathrm{mass}}[\mathbf Y].
\end{equation}
With \eqref{eq:mass_correction}, these equations have the common effective adjacent coupling $i/(\sqrt2\lambda)$. Vanishing mixed divergence makes the unsymmetrized adjacent derivatives symmetric in all free indices of the corresponding spinor type. In two-dimensional spin spaces, this is equivalent to
\begin{equation}
\partial_B{}^{B'}Y^{(2s-k,k)}{}_{B'A'_2\cdots A'_k}{}^{BA_{k+2}\cdots A_{2s}}=0,
\qquad 1\leq k<2s.
\end{equation}
The symmetric projection is retained, so the free system does not require $\mathbf J=\mathbf J_{\mathrm{mass}}[\mathbf Y]$. Together, these equations give the \textit{Dirac--Fierz--Pauli system} in symmetric multispinor form \cite{Dirac1936,FierzPauli1939,BargmannWigner1948}. For higher spin, its extension to a general curved background requires more than the replacement $\partial\mapsto\nabla$: curvature commutators introduce compatibility conditions that can restrict the background and admissible fields \cite{Buchdahl1958}.

For $m\neq0$, either extremal component of a solution determines the multiplet:
\begin{align}
\partial_{A_1A'_1}\cdots\partial_{A_kA'_k}
Y^{(2s,0)\,A_1\cdots A_{2s}}
&=\left(-\frac{i}{\sqrt2\lambda}\right)^k
Y^{(2s-k,k)}{}_{A'_1\cdots A'_k}{}^{A_{k+1}\cdots A_{2s}},\\
\partial^{A_{k+1}A'_{k+1}}\cdots\partial^{A_{2s}A'_{2s}}
Y^{(0,2s)}{}_{A'_1\cdots A'_{2s}}
&=\left(-\frac{i}{\sqrt2\lambda}\right)^{2s-k}
Y^{(2s-k,k)}{}_{A'_1\cdots A'_k}{}^{A_{k+1}\cdots A_{2s}},
\end{align}
where $0\leq k\leq2s$, with empty products understood as the identity. The complete first-order system implies
\begin{equation}
(\Box+m^2)Y^{(2s-k,k)}=0,\qquad
\Box:=\partial^{AA'}\partial_{AA'}.
\end{equation}
Thus all components have the same mass; $k$ labels their chiral spinor representations.

On the full flat background, sufficiently regular classical solutions admit the expansion
\begin{equation}\label{eq:Y_gen}
\mathbf Y(x)=
\sum_{r=-s}^{s}
\int\frac{d^3\mathbf q}{(2\pi)^3\,2E_{\mathbf q}}
\left[
a_r(\mathbf q)\mathbf y_{+,r}(q)e^{-iq\cdot x}
+b_r(\mathbf q)\mathbf y_{-,r}(q)e^{iq\cdot x}
\right],
\qquad
E_{\mathbf q}=\sqrt{|\mathbf q|^2+m^2}.
\end{equation}
For $q^2=m^2$ and $q^0>0$, the component polarizations satisfy
\begin{equation}\label{eq:ladder_mode}
y_{\pm,r}^{(2s-k,k)}{}_{A'_1\cdots A'_k}{}^{A_{k+1}\cdots A_{2s}}
=(\pm\sqrt2\lambda)^k
q_{A_1A'_1}\cdots q_{A_kA'_k}
y_{\pm,r}^{(2s,0)\,A_1\cdots A_{2s}}.
\end{equation}
At each momentum and frequency sign, the $2s+1$ polarizations carry the massive spin-$s$ representation of the little group \cite{BargmannWigner1948,Weinberg1995}. The amplitudes are classical; their quantization is discussed in Section~\ref{sec:interpretation_tt}. The scalar case uses $\mathcal T_Y^{(0)}=U\times\mathbb C$ and $(\Box+m^2)Y^{(0,0)}=0$, with a single polarization. Its second-order equation is specified separately because the rank-zero bundle has no spinor contractions.

\paragraph*{\textbf{Massless Fields from General Chiral Twinors}}
For $s>0$, the free massless realization on $\mathbf W\in\Gamma(U,\mathcal T_W^{(2s)})$ is specified by the zero-rest-mass equations \cite{PenroseRindler1984,Penrose1969}
\begin{equation}\label{eq:ZRM}
\partial_{A_1B'}W^{(2s,0)\,A_1\cdots A_{2s}}=0,\qquad
\partial^{BA'_1}W^{(0,2s)}{}_{A'_1\cdots A'_{2s}}=0.
\end{equation}
This realization uses the chiral connection without the particle soldering coupling or the mass-normalization source. On a Lorentzian real slice, these describe helicities $\pm s$; selecting the extremal sectors is additional to taking $\lambda^{-1}\to0$ in the full ladder multiplet.

For $s\geq1$, the encoding \eqref{eq:chiral_twinor_curvature} makes the free equations \eqref{eq:ZRM} equivalent to $dF_L=dF_R=0$ in the parallel frame. On a contractible patch, this gives the linear potential description
\begin{equation}
F_{L,R}=dA_{L,R},\qquad
\delta A_{L,R}=d\chi_{L,R},
\end{equation}
where the coefficient bundles are $\bigodot^{2s-2}S^A$ and $\bigodot^{2s-2}S_{A'}$, respectively. The chiral and symmetry conditions on the field strengths encode the free equations.

Beyond the free theory, the spin-$1$ sector admits a nonlinear gauge realization with an internal gauge group $K$:
\begin{equation}\label{eq:potential_to_twinor}
\begin{gathered}
\mathbf W\in\Gamma(U,\mathcal T_W^{(2)}\otimes\operatorname{ad}Q),\qquad
F=(\iota_1\otimes\mathbf1)\mathbf W=dA+A\wedge A,\\
A\longmapsto g^{-1}Ag+g^{-1}dg,\qquad
F\longmapsto g^{-1}Fg,\qquad
\mathbf W\longmapsto\operatorname{Ad}_{g^{-1}}\mathbf W,
\end{gathered}
\end{equation}
where $g:U\to K$. With $\mathcal D_{AA'}:=\nabla_{AA'}+[A_{AA'},\,\cdot\,]$, the source-free equations are
\begin{equation}
\mathcal D_{AB'}W^{(2,0)\,AB}=0,\qquad
\mathcal D^{BA'}W^{(0,2)}{}_{A'B'}=0
\qquad\Longleftrightarrow\qquad
D^{\mathrm{ad}}F=0,\qquad
D^{\mathrm{ad}}\star F=0.
\end{equation}
These are the Bianchi identity and Yang--Mills equation in spinor form \cite{PenroseRindler1984,YangMills1954}; they also hold on torsion-free curved backgrounds with the Levi--Civita spin connection. The Hodge operator is defined by $\alpha\wedge\star\beta=\langle\alpha,\beta\rangle_g\,\mathrm{vol}_g$; in Lorentzian signature, $\star^2=-1$ on two-forms. On a physical real slice, choosing the internal real form $U(1)$ gives vacuum Maxwell theory, while $SU(3)$ gives the classical pure gluon sector of QCD. The field strength must also satisfy the corresponding reality condition, relating its two chiral components. The same potential induces the internal connection acting on matter in any specified representation.

Explicitly, in a local gauge with a basis $t_I$ of $\mathfrak k$, the spin-one field strength is
\begin{equation}
F=\frac12\left(
\epsilon_{A'B'}W^{(2,0)I}{}_{AB}
+\epsilon_{AB}W^{(0,2)I}{}_{A'B'}
\right)t_I\,e^{AA'}\wedge e^{BB'}.
\end{equation}
The index $I$ labels the internal gauge components, while the symmetric spinor indices describe the two chiral components of the field strength.

\paragraph*{\textbf{Dirac--Maxwell--Yukawa Example}}
For a test-field example on a fixed flat, torsion-free background, Lorentzian reality and the internal real form $U(1)$ are imposed. The particle scale lies on the positive-mass AdS branch, $m=\lambda^{-1}>0$, as in Section~\ref{sec:freefield_spin}. For $s=\tfrac12$, the mass-normalization source vanishes and no mixed contraction is present. With $\mathcal D_{AA'}:=\partial_{AA'}+iqA_{AA'}$, the equations are
\begin{equation}
\mathcal D_{AA'}u^A+\frac{im}{\sqrt2}v_{A'}=j_{A'},\qquad
\mathcal D^{AA'}v_{A'}+\frac{im}{\sqrt2}u^A=\widetilde j^A.
\end{equation}
Minimal electromagnetic coupling is already included in $\mathcal D$ and requires no additional contracted source. A real neutral scalar $\phi$ with real Yukawa coupling $g_Y$ contributes
\begin{equation}
j_{A'}=-\frac{ig_Y}{\sqrt2}\phi v_{A'},\qquad
\widetilde j^A=-\frac{ig_Y}{\sqrt2}\phi u^A,
\end{equation}
equivalently replacing $m$ by $m+g_Y\phi$ \cite{PeskinSchroeder1995}. The accompanying gauge and scalar equations on this flat background are
\begin{equation}
\begin{gathered}
F=dA,\qquad
d\star F=-\star j_{\mathrm{em}},\qquad
(j_{\mathrm{em}})_{AA'}=q\overline{Y_d}\gamma_{AA'}Y_d,\\
\Box\phi+V'(\phi)=-g_Y\overline{Y_d}Y_d.
\end{gathered}
\end{equation}
Here $\overline{Y_d}$ denotes the Dirac adjoint and the scalar has a canonical kinetic term. With the stated Hodge convention, the Maxwell equation is equivalent to $\partial_aF^{ab}=j_{\mathrm{em}}^b$. For real couplings, charge conservation follows from the Dirac equation and its adjoint. The scalar contributions are omitted in the Dirac--Maxwell system. Coupling this example to Einstein--Cartan gravity requires relaxing the fixed flat-background assumption and using the full spin connection, or the equivalent contorsion contribution in \eqref{eq:contorsion_matter_source}.

The contracted spinor sources, electric current, energy--momentum tensor, and spin current are distinct. In particular, $j_{A'}=\widetilde j^A=0$ does not imply vanishing charge, energy--momentum, or spin. Before restriction to the real slice, the holomorphic matter and dual matter fields are independent.

\subsection{Nonlinear Gravity and Adjoint Twinor Deformations}\label{sec:adjoint_twinor}

The full adjoint Cartan curvature $\mathcal F$ satisfies the Bianchi identity \eqref{eq:adjoint_bianchi} independently of the gravitational field equations. Using the chiral grading $\Upsilon$ introduced above, define the soldering matrix by
\begin{equation}
\mathbf e:=
\begin{pmatrix}
0&e^{AB'}\\
e_{A'B}&0
\end{pmatrix}.
\end{equation}
Here the orientation is fixed so that \mbox{$\star e^{AA'}:=\frac i3 e^{AC'}\wedge e_{C'B}\wedge e^{BA'}$}.

Einstein--Cartan dynamics is specified by the curvature--soldering source equation
\begin{equation}\label{eq:nonlinear_adjoint_gravity}
\mathcal F\wedge\mathbf e+\mathbf e\wedge\Upsilon\mathcal F\Upsilon
=
8\pi Gi
\begin{pmatrix}
\dfrac1{\sqrt2\ell}\mathcal S_L{}^A{}_B
&
\mathcal J_T^{AB'}\\[2mm]
-\mathcal J_{T\,A'B}
&
\dfrac1{\sqrt2\ell}\mathcal S_R{}_{A'}{}^{B'}
\end{pmatrix}.
\end{equation}
Here $\mathcal J_T^{AA'}:=T_{BB'}{}^{AA'}\star e^{BB'}$ is the energy--momentum three-form, with $T_{AA'BB'}$ the tetrad energy--momentum tensor; $\mathcal J_{T\,A'B}$ denotes its spinor-lowered form. The three-forms $\mathcal S_L,\mathcal S_R$ are the chiral Lorentz spin currents. Conjugation by $\Upsilon$ preserves the diagonal curvature blocks and reverses the off-diagonal torsion blocks. Since $\Upsilon$ commutes with $H$, this is an $H$-equivariant equation of endomorphism-valued three-forms, defined before any reality condition is imposed.

Its independent blocks give
\begin{equation}
\begin{gathered}
R_L{}^A{}_B\wedge e^{BA'}
+e^{AB'}\wedge R_R{}_{B'}{}^{A'}
+\frac{\Lambda}{3}e^{AC'}\wedge e_{C'B}\wedge e^{BA'}
=8\pi Gi\,T_{BB'}{}^{AA'}\star e^{BB'},\\
\tau^{AC'}\wedge e_{C'B}-e^{AC'}\wedge\tau_{C'B}
=8\pi G\,\mathcal S_L{}^A{}_B,\qquad
\tau_{A'C}\wedge e^{CB'}-e_{A'C}\wedge\tau^{CB'}
=8\pi G\,\mathcal S_R{}_{A'}{}^{B'}.
\end{gathered}
\end{equation}
The cosmological term, with $\Lambda=-3/\ell^2$, arises directly from the diagonal soldering contributions in $\mathcal F$. The lower-left block is the negative spinor-lowered upper-right block, so it supplies no additional equation. The source normalizations are fixed by \eqref{eq:nonlinear_adjoint_gravity}, while their field dependence is supplied by the matter theory. For nondegenerate soldering, the spin-source equations determine torsion algebraically.

The source equation is imposed directly, independently of the action formulations above. Its compatibility is constrained by the adjoint Bianchi identity, whose off-diagonal component fixes the complementary difference
\begin{equation}
R_L{}^A{}_B\wedge e^{BA'}
-e^{AB'}\wedge R_R{}_{B'}{}^{A'}
=d^\omega\tau^{AA'}.
\end{equation}
With the adopted curvature convention, the upper-right block of \eqref{eq:nonlinear_adjoint_gravity} is equivalent to\footnote{$i\left(R_L{}^A{}_B\wedge e^{BA'}+e^{AB'}\wedge R_R{}_{B'}{}^{A'}\right)=\left(R_{BB'}{}^{AA'}-\frac12R\,\delta^A{}_B\delta^{A'}{}_{B'}\right)\star e^{BB'}$.}
\begin{equation}\label{eq:nonlinear_spinor_einstein}
R_{AA'BB'}-\frac12R\,\epsilon_{AB}\epsilon_{A'B'}
+\Lambda\epsilon_{AB}\epsilon_{A'B'}
=-8\pi G\,T_{AA'BB'}.
\end{equation}
The Ricci tensor is formed from the full metric-compatible connection and need not be symmetric in the presence of torsion.

\paragraph*{\textbf{Torsion-Free, Vacuum, and Flat Reductions}}
Vanishing spin currents imply vanishing torsion, reducing \eqref{eq:nonlinear_spinor_einstein} to the Levi--Civita Einstein equation with $\nabla^{AA'}T_{AA'BB'}=0$. In vacuum, $R_{AA'BB'}=\Lambda\epsilon_{AB}\epsilon_{A'B'}$, and the remaining Cartan curvature is encoded by the Weyl chiral twinor in $\mathcal T_W^{(4)}$. Its covariant zero-rest-mass equations follow from the Bianchi identity, with the connection determined simultaneously by the gravitational equations.

Cartan flatness, $\mathcal F=0$, additionally removes the Weyl curvature and gives the local homogeneous model. By contrast, $R_L=R_R=0$ and $\tau=0$ give a flat metric, requiring $T_{AA'BB'}=-\Lambda(8\pi G)^{-1}\epsilon_{AB}\epsilon_{A'B'}$. On a Lorentzian real slice, the corresponding reality conditions are imposed on the soldering, connections, and currents.

\paragraph*{\textbf{Adjoint BGG and Gravitational Perturbations}}
For a reference Cartan connection $\mathcal A_0$, a finite adjoint-valued deformation $\alpha$ gives
\begin{equation}\label{eq:finite_adjoint_deformation}
\mathcal A=\mathcal A_0+\alpha,\qquad
\mathcal F=\mathcal F_0+d^{\nabla^{\mathcal A_0}}\alpha+\alpha\wedge\alpha,
\end{equation}
provided the deformed soldering remains nondegenerate. Substitution into \eqref{eq:nonlinear_adjoint_gravity}, together with the deformed soldering matrix, retains the full nonlinear equations.

Torsion-free infinitesimal deformations of the reference geometry are organized by \eqref{eq:full-BGG-adjoint}. With $\delta g=2\psi$, the symmetric coframe gauge gives $\delta e^{AA'}=\psi^{AA'}{}_{BB'}e^{BB'}$, and the connection variation is determined by the linearized torsion equation. On a torsion-free vacuum Einstein background, with fixed $\Lambda$ and vanishing linearized spin source, the Einstein projection is
\begin{equation}\label{eq:adjoint_einstein_projection}
\delta\mathring G_{AA'BB'}[2\psi]+2\Lambda\psi_{AA'BB'}
=-8\pi G\,\delta T_{AA'BB'},\qquad
\mathring\nabla^{AA'}\delta T_{AA'BB'}=0.
\end{equation}
Here $\mathring G$ is the Levi--Civita Einstein tensor, and $\delta\mathring G[2\psi]=\left.\frac{d}{dt}\right|_{t=0}\mathring G(g(t))$ for $g(0)=g$ and $\dot g(0)=2\psi$. Tensor variations are expressed using the background frame identification. The curvature deformation represented by $D^1\psi$ enters together with the variation of the soldering. Nonzero background Weyl curvature is allowed: source conservation follows from the linearized Bianchi identity, independently of whether the BGG sequence is a complex. On backgrounds with matter, the variations of the background sources and their conservation equations must also be retained.

\subsection{Twinor Field Equation and its Projections on Curved Backgrounds}\label{sec:sourced_twinor_transport}

\paragraph*{\textbf{Gravity and Torsion}}
For nondegenerate soldering, contorsion is determined algebraically by the total spin current through \eqref{eq:nonlinear_adjoint_gravity}. Writing $\omega_{L,R}=\mathring\omega_{L,R}+\kappa_{L,R}$, where $\mathring\omega_{L,R}$ are the Levi--Civita spin connections, gives
\begin{equation}
\begin{aligned}
R_{L,R}
&=\mathring R_{L,R}
+d^{\mathring\omega}\kappa_{L,R}
+\kappa_{L,R}\wedge\kappa_{L,R},\\
\tau^{AA'}
&=\kappa_L{}^A{}_B\wedge e^{BA'}
+e^{AB'}\wedge\kappa_R{}_{B'}{}^{A'}.
\end{aligned}
\end{equation}
Let $\mathring D$ denote the coupled twinor derivative obtained by replacing $\omega_{L,R}$ with $\mathring\omega_{L,R}$ while retaining the same soldering, particle scale, and internal connection. Then
\begin{equation}
D\mathbf Y=\mathring D\mathbf Y+
\rho_{2s*}
\begin{pmatrix}
\kappa_L&0\\
0&\kappa_R
\end{pmatrix}
\mathbf Y,
\end{equation}
where $\rho_{2s*}$ acts on each symmetric twinor slot. The final term is a one-form, algebraic in contorsion and $\mathbf Y$. The equations for the constrained projections can therefore be written as
\begin{equation}\label{eq:contorsion_matter_source}
\begin{aligned}
\operatorname{ctr}(\mathring D\mathbf Y)
&=\mathbf j_{\mathrm{mass}}[\mathbf Y]+\mathbf j_{\mathrm{phys}}
-\operatorname{ctr}\left[
\rho_{2s*}
\begin{pmatrix}
\kappa_L&0\\
0&\kappa_R
\end{pmatrix}
\mathbf Y
\right],\\
\operatorname{ctr}_0(\mathring D\mathbf Y)
&=\mathbf j_0
-\operatorname{ctr}_0\left[
\rho_{2s*}
\begin{pmatrix}
\kappa_L&0\\
0&\kappa_R
\end{pmatrix}
\mathbf Y
\right].
\end{aligned}
\end{equation}
The contractions and mass-normalization source use the same soldering form and particle scale. The mixed equation is absent for a Dirac twinor. These are exact rearrangements. Eliminating contorsion further requires substituting its algebraic solution from the spin-source equations.

\paragraph*{\textbf{Curvature Compatibility}}
The compatibility identity \eqref{eq:master_twinor_equation} applies to the total source, including its mass-normalization and symmetric derivative contributions:
\begin{equation}\label{eq:twinor_curvature_compatibility}
D\mathbf J=D^2\mathbf Y=\mathcal F_D\mathbf Y.
\end{equation}
This identity remains valid with torsion. The curvature acts on each twinor slot and on the internal coefficient bundle:
\begin{equation}
(\mathcal F_D\mathbf Y)^{\alpha_1\cdots\alpha_{2s}}
=\sum_{r=1}^{2s}(\mathcal F_\lambda)^{\alpha_r}{}_\beta
Y^{\alpha_1\cdots\alpha_{r-1}\beta\alpha_{r+1}\cdots\alpha_{2s}}
+F_V\mathbf Y^{\alpha_1\cdots\alpha_{2s}}.
\end{equation}
These contributions are given by \eqref{eq:coupled_twinor_connection} and \eqref{eq:particle_twinor_curvature}. The spin curvature and torsion are constrained by the gravitational source equations, the particle soldering contribution depends on $\lambda$, and the internal curvature is governed by the gauge-field equations.

An explicit compatibility condition is obtained from the first two adjacent equations. On a torsion-free background without internal gauge fields, with constant $\lambda\neq0$ and $2s\geq2$, subtraction of the mass-normalization source gives
\begin{equation}\label{eq:ladder_first_two_sourced}
\begin{aligned}
\nabla_{BA'}Y^{(2s,0)\,BC A_3\cdots A_{2s}}
&=-\frac{i}{\sqrt2\lambda}Y^{(2s-1,1)}{}_{A'}{}^{C A_3\cdots A_{2s}}
+j_{L,0\,A'}{}^{C A_3\cdots A_{2s}},\\
\nabla_{CB'}Y^{(2s-1,1)}{}_{A'}{}^{C A_3\cdots A_{2s}}
&=-\frac{i}{\sqrt2\lambda}Y^{(2s-2,2)}{}_{B'A'}{}^{A_3\cdots A_{2s}}
+j_{L,1\,B'A'}{}^{A_3\cdots A_{2s}}
+P_{B'A'}{}^{A_3\cdots A_{2s}}.
\end{aligned}
\end{equation}
The adjacent equation determines the symmetric part of the second line. Its antisymmetric part is fixed by the mixed source:
\begin{equation}\label{eq:ladder_mixed_source_antisymmetric}
P_{B'A'}{}^{A_3\cdots A_{2s}}
:=\nabla_{C[B'}Y^{(2s-1,1)}{}_{A']}{}^{C A_3\cdots A_{2s}}
=-\frac12\epsilon_{B'A'}(\mathbf j_0)_1{}^{A_3\cdots A_{2s}},
\end{equation}
where square brackets denote unit-weight antisymmetrization and $(\mathbf j_0)_1$ is the mixed source of the $(2s-1,1)$ component. Thus $P$ introduces no independent field.

Applying $\nabla_{CB'}$ to the first line of \eqref{eq:ladder_first_two_sourced}, substituting the second, and antisymmetrizing in $B',A'$ gives
\begin{equation}\label{eq:ladder_source_curvature_condition}
[\nabla_{CB'},\nabla_{BA'}]Y^{(2s,0)\,BC A_3\cdots A_{2s}}
=\nabla_{CB'}j_{L,0\,A'}{}^{C A_3\cdots A_{2s}}
-\nabla_{CA'}j_{L,0\,B'}{}^{C A_3\cdots A_{2s}}
-\frac{\sqrt2 i}{\lambda}P_{B'A'}{}^{A_3\cdots A_{2s}}.
\end{equation}
The symmetric mass term and $j_{L,1}$ cancel, while symmetry in the contracted unprimed indices converts the second derivatives into the displayed commutator. The curvature action is therefore balanced by derivatives of the adjacent source and by the mixed source. When $\mathbf j_{\mathrm{phys}}=\mathbf j_0=0$, this condition reduces to
\begin{equation}
[\nabla_{CB'},\nabla_{BA'}]Y^{(2s,0)\,BC A_3\cdots A_{2s}}=0.
\end{equation}
This is one necessary compatibility condition of the constrained system; its validity on a curved background is not ensured by retaining the symmetric derivative projection. Compatibility must be checked for the specified source relations and accompanying field equations.

\subsection{Physical Interpretation of Twinor Theory}\label{sec:interpretation_tt}

\paragraph*{\textbf{Twinor Field Equation and Projection Constraints}}
The source relations in \eqref{eq:total_twinor_source} determine the physical content of $D\mathbf Y=\mathbf J$. Vanishing additional contracted sources leaves the minimal gravitational and gauge couplings in $D$ intact and does not require the total source one-form to vanish.

Together with the energy--momentum and spin-source equations \eqref{eq:nonlinear_adjoint_gravity} and the equations for all dynamical matter and gauge fields, these relations define the coupled system. Its compatibility must be checked for the chosen couplings.

\paragraph*{\textbf{Imaginary Coupling and the Anti-de Sitter Reality Condition}}
With the mass normalization in \eqref{eq:mass_correction}, the massive equations imply $m^2=\lambda^{-2}$. Positive real mass squared therefore requires $\lambda\in\mathbb R\setminus\{0\}$, the anti-de Sitter reality condition for the scale-deformed particle connection. On the positive-mass branch, the induced soldering coupling together with the mass-normalization source gives
\begin{equation}
\frac{i}{\sqrt2\lambda}=\frac{im}{\sqrt2},
\qquad m=\lambda^{-1}>0.
\end{equation}
The imaginary factor is already present in the classical soldering coupling and is retained by the mass-normalization source. The geometric scale $\ell$, with $\Lambda=-3/\ell^2$, is independent of $\lambda$; the particle reality condition does not select the cosmological constant or require a locally anti-de Sitter metric. Physical Dirac fields remain complex unless an additional Majorana condition is imposed. Neither quantum commutation relations nor a quantum state follows from the coupling phase.

\paragraph*{\textbf{Quantization}}
For free fields on the flat background, the amplitudes in \eqref{eq:Y_gen} are promoted to annihilation and creation operators, $a_r\mapsto\widehat a_r$ and $b_r\mapsto\widehat b_r^\dagger$, with commutators for integer spin and anticommutators for half-integer spin \cite{Weinberg1995}. The linear field equations are retained. On curved backgrounds, quantization is instead formulated through the field algebra and its states, without assuming a preferred positive-frequency decomposition \cite{Wald1994}.

The adjoint BGG description contributes the classical gauge and curvature operators relevant to gravitational perturbations. On a torsion-free vacuum Einstein background, define the linearized Einstein operator in the normalization $\delta g=2\psi$ by
\begin{equation}\label{eq:quantization_linearized_einstein}
\mathcal P_g\psi:=\delta\mathring G[2\psi]+2\Lambda\psi,
\qquad
\mathcal P_gD^0\xi=0,
\qquad
\psi\sim\psi+D^0\xi.
\end{equation}
The classical solution space modulo gauge is $\ker\mathcal P_g/\operatorname{im}D^0$, with compatible support and boundary conditions. The BGG operator $D^1$ describes curvature deformation, while $\mathcal P_g$ imposes the linearized Einstein equation. The identity $\mathcal P_gD^0=0$ holds on vacuum Einstein backgrounds even when the BGG sequence is not a complex.

Quantization requires a compatible presymplectic structure, causal propagators after suitable gauge fixing, and positive states with the appropriate Hadamard singularity structure \cite{FewsterHunt2013,HackSchenkel2013}. Globally hyperbolic backgrounds provide the standard setting; timelike boundaries require suitable boundary conditions. The BGG operators organize classical gauge and curvature data but do not supply these quantum structures. Their application to twinor quantization remains a direction for further work.

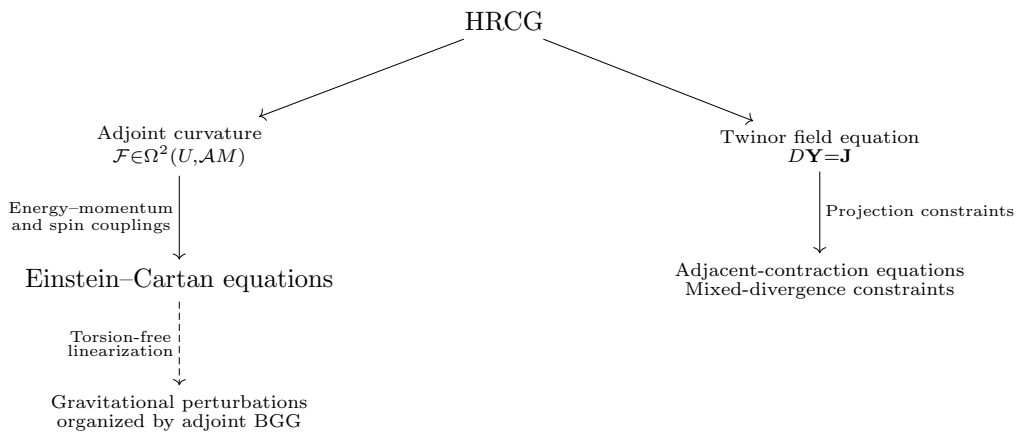
\begin{figure}[htbp]
\centering
\begin{tikzcd}[
    row sep=3.2em,
    column sep=4.5em,
    cells={font=\normalsize},
    every arrow/.append style={font=\small}
]
& \text{HRCG}
    \ar[dl]
    \ar[dr]
& \\
\substack{\text{Adjoint curvature}\\
          \mathcal F\in\Omega^2(U,\mathcal AM)}
    \ar[d,"\substack{\text{Energy--momentum}\\
                    \text{and spin couplings}}"']
& &
\substack{\text{Twinor field equation}\\
          D\mathbf Y=\mathbf J}
    \ar[d,"\text{Projection constraints}"]
\\
\text{Einstein--Cartan equations}
    \ar[d,dashed,
        "\substack{\text{Torsion-free}\\
                   \text{linearization}}"']
& &
\substack{\text{Adjacent-contraction equations}\\
          \text{Mixed-divergence constraints}}
\\
\substack{\text{Gravitational perturbations}\\
          \text{organized by adjoint BGG}}
& &
\end{tikzcd}
\caption{Holomorphic reductive Cartan geometry supplies the adjoint curvature and the coupled twinor connections. The twinor field equation is resolved into adjacent-contraction equations, mixed-divergence constraints, and a symmetric projection with no independently imposed field equation. Source couplings specify the dynamics, while the Bianchi and curvature identities constrain compatibility. The adjoint BGG sequence organizes torsion-free infinitesimal deformations; reality conditions select the physical realization.}
\label{fig:twinor_framework}
\end{figure}

\section{Conclusion}

Holomorphic reductive Cartan geometry of type $(\mathrm{Sp}(4,\mathbb C),\mathrm{SL}(2,\mathbb C)_L\times\mathrm{SL}(2,\mathbb C)_R)$ provides a common spacetime framework for Dirac, ladder, and chiral twinors. Their dynamics is expressed through the twinor field equation $D\mathbf Y=\mathbf J$ and its projection constraints, with curvature compatibility given by $D\mathbf J=\mathcal F_D\mathbf Y$. On a flat torsion-free Lorentzian background, the adjacent-contraction equations with the mass-normalization source and vanishing additional contracted sources, together with the mixed-divergence constraints, reproduce the massive Dirac--Fierz--Pauli system. The symmetric projection retains the remaining derivative data. The chiral realization gives the zero-rest-mass equations; with the corresponding gauge-curvature relations and gauge-covariant derivatives, the spin-one sector gives Maxwell and Yang--Mills theory.

The full adjoint curvature encodes nonlinear Einstein--Cartan gravity, with energy--momentum and spin currents sourcing its curvature and torsion equations. These equations are imposed directly, while the adjoint Bianchi identity constrains their compatibility. The adjoint BGG-type sequence organizes torsion-free infinitesimal geometry, and the linearized Einstein equation selects gravitational perturbations. This description applies on curved Einstein backgrounds without requiring the BGG sequence to be a complex.

A central feature is the correspondence, within the fixed mass normalization, between positive real mass squared and the anti-de Sitter reality condition on the particle scale. The imaginary mass coupling is therefore present classically, before quantization. The independence of the particle scale $\lambda$ and cosmological scale $\ell$ allows this condition without requiring an anti-de Sitter physical metric.

The relation between fields and geometry may therefore be summarized as
\begin{quote}
\textbf{\textit{Fields source spacetime geometry; the twinor field equation expresses their dynamics.}}
\end{quote}
The bundle construction extends to arbitrary spin; compatibility of particular coupled field systems depends on their source relations and dynamical equations. Further work may investigate their role in constructing gauge-invariant observables and quantization on curved backgrounds. Causal propagation, a compatible field algebra, and admissible quantum states remain additional structures.

\begin{acknowledgments}
I thank Tim Adamo and Lionel Mason for discussions at the very beginning of this project, during which we developed the initial idea that ultimately took the form presented here. I also thank the other participants in the Simons Programme on Twistor Theory and its Applications for several helpful discussions, and Jan Slov\'ak for discussions throughout this work and for reviewing the manuscript.

The author was supported by the Czech Science Foundation (GA\v{C}R) under the research project ``Cartan supergeometries and Higher Cartan geometries'' (Grant No.~GA24-10887S) and by the European Union's Horizon Europe research and innovation programme under the Marie Sk\l{}odowska-Curie grant agreement No.~101086123 (CaLIGOLA). This research was supported by the Simons Foundation grant SFI-MPS-T-Institutes-00010825, and by State Treasury funds as part of a task commissioned by the Minister of Science and Higher Education under the project ``Organization of the Simons Semesters at the Banach Center --- New Energies in 2026--2028'' (MNiSW/2025/DAP/491).
\end{acknowledgments}


\begin{thebibliography}{99}
\interlinepenalty=10000


\bibitem{Penrose1967}
R.~Penrose,
\textit{Twistor algebra},
J. Math. Phys. \textbf{8} (1967) 345--366.

\bibitem{PenroseMacCallum1973}
R.~Penrose and M.~A.~H.~MacCallum,
\textit{Twistor theory: An approach to the quantisation of fields and space-time},
Phys. Rep. \textbf{6} (1973) 241--315,
doi:10.1016/0370-1573(73)90008-2.

\bibitem{PenroseRindler1986}
R.~Penrose and W.~Rindler,
\textit{Spinors and Space-Time, Volume 2: Spinor and Twistor Methods in Space-Time Geometry},
Cambridge University Press, Cambridge (1986).

\bibitem{BastonEastwood1989}
R.~J.~Baston and M.~G.~Eastwood,
\textit{The Penrose Transform: Its Interaction with Representation Theory},
Oxford Mathematical Monographs, Clarendon Press, Oxford (1989).

\bibitem{Penrose1976}
R.~Penrose,
\textit{Nonlinear gravitons and curved twistor theory},
Gen. Relativ. Gravit. \textbf{7} (1976) 31--52.

\bibitem{Ward1977}
R.~S.~Ward,
\textit{On self-dual gauge fields},
Phys. Lett. A \textbf{61} (1977) 81--82.

\bibitem{Penrose2015}
R.~Penrose,
\textit{Palatial twistor theory and the twistor googly problem},
Philos. Trans. R. Soc. A \textbf{373} (2015) 20140237,
doi:10.1098/rsta.2014.0237.

\bibitem{Isenberg1978}
J.~Isenberg, P.~B.~Yasskin and P.~S.~Green,
\textit{Non-self-dual gauge fields},
Phys. Lett. B \textbf{78} (1978) 462--464.

\bibitem{Witten1978}
E.~Witten,
\textit{An interpretation of classical Yang--Mills theory},
Phys. Lett. B \textbf{77} (1978) 394--398.

\bibitem{Mason2005}
L.~J.~Mason,
\textit{Twistor actions for non-self-dual fields; a new foundation for twistor-string theory},
J. High Energy Phys. \textbf{10} (2005) 009,
arXiv:hep-th/0507269.

\bibitem{Tod1977}
K.~P.~Tod,
\textit{Some symplectic forms arising in twistor theory},
Rep. Math. Phys. \textbf{11} (1977) 339--346.

\bibitem{Sharpe1997}
R.~W.~Sharpe,
\textit{Differential Geometry: Cartan's Generalization of Klein's Erlangen Program},
Graduate Texts in Mathematics \textbf{166}, Springer, New York (1997).

\bibitem{BaileyEastwoodGover1994}
T.~N.~Bailey, M.~G.~Eastwood and A.~R.~Gover,
\textit{Thomas's structure bundle for conformal, projective and related structures},
Rocky Mountain J. Math. \textbf{24} (1994) 1191--1217.

\bibitem{AttardFrancois2016_1}
J.~Attard and J.~Fran\c{c}ois,
\textit{Tractors and twistors from conformal Cartan geometry: A gauge theoretic approach I. Tractors},
Adv. Theor. Math. Phys. \textbf{22} (2018) 1831--1883,
arXiv:1609.07307 [math-ph].

\bibitem{AttardFrancois2016_2}
J.~Fran\c{c}ois and J.~Attard,
\textit{Tractors and twistors from conformal Cartan geometry: A gauge theoretic approach II. Twistors},
Class. Quantum Grav. \textbf{34} (2017) 085004,
arXiv:1611.03891 [math-ph].

\bibitem{CapSlovakSoucek2001}
A.~\v{C}ap, J.~Slov\'ak and V.~Sou\v{c}ek,
\textit{Bernstein--Gelfand--Gelfand sequences},
Ann. of Math. \textbf{154} (2001) 97--113,
arXiv:math/0001164.

\bibitem{Chien2026}
C.-H.~Chien,
\textit{Holomorphic Conformal Cartan Geometry: Conformal Gravity, Twistors, and Ambitwistors},
arXiv:2609.22861 [math-ph] (2026).

\bibitem{Dirac1936}
P.~A.~M.~Dirac,
\textit{Relativistic wave equations},
Proc. R. Soc. Lond. A \textbf{155} (1936) 447--459.

\bibitem{FierzPauli1939}
M.~Fierz and W.~Pauli,
\textit{On relativistic wave equations for particles of arbitrary spin in an electromagnetic field},
Proc. R. Soc. Lond. A \textbf{173} (1939) 211--232.

\bibitem{BargmannWigner1948}
V.~Bargmann and E.~P.~Wigner,
\textit{Group theoretical discussion of relativistic wave equations},
Proc. Natl. Acad. Sci. U.S.A. \textbf{34} (1948) 211--223.

\bibitem{Cap2026}
A.~\v{C}ap,
\textit{A new construction of the Riemannian deformation sequence},
arXiv:2607.26653v1 [math.DG] (2026).

\bibitem{MacDowellMansouri1977}
S.~W.~MacDowell and F.~Mansouri,
\textit{Unified geometric theory of gravity and supergravity},
Phys. Rev. Lett. \textbf{38} (1977) 739--742;
erratum, Phys. Rev. Lett. \textbf{38} (1977) 1376.

\bibitem{FreidelStarodubtsev2005}
L.~Freidel and A.~Starodubtsev,
\textit{Quantum gravity in terms of topological observables},
arXiv:hep-th/0501191 (2005).

\bibitem{Buchdahl1958}
H.~A.~Buchdahl,
\textit{On the compatibility of relativistic wave equations for particles of higher spin in the presence of a gravitational field},
Nuovo Cimento \textbf{10} (1958) 96--103.

\bibitem{PenroseRindler1984}
R.~Penrose and W.~Rindler,
\textit{Spinors and Space-Time, Volume 1: Two-Spinor Calculus and Relativistic Fields},
Cambridge University Press, Cambridge (1984).

\bibitem{Plebanski1977}
J.~F.~Pleba\'nski,
\textit{On the Separation of Einsteinian Substructures},
J. Math. Phys. \textbf{18} (1977) no.~12, 2511--2520,
doi:10.1063/1.523215.

\bibitem{Ashtekar1986}
A.~Ashtekar,
\textit{New Variables for Classical and Quantum Gravity},
Phys. Rev. Lett. \textbf{57} (1986) no.~18, 2244--2247,
doi:10.1103/PhysRevLett.57.2244.

\bibitem{Barbero1995}
J.~F.~Barbero G.,
\textit{Real Ashtekar Variables for Lorentzian Signature Space-Times},
Phys. Rev. D \textbf{51} (1995) 5507--5510,
arXiv:gr-qc/9410014.

\bibitem{Holst1996}
S.~Holst,
\textit{Barbero's Hamiltonian Derived from a Generalized Hilbert-Palatini Action},
Phys. Rev. D \textbf{53} (1996) 5966--5969,
arXiv:gr-qc/9511026.

\bibitem{HehlEtAl1976}
F.~W.~Hehl, P.~von der Heyde, G.~D.~Kerlick, and J.~M.~Nester,
\textit{General Relativity with Spin and Torsion: Foundations and Prospects},
Rev. Mod. Phys. \textbf{48} (1976) no.~3, 393--416,
doi:10.1103/RevModPhys.48.393.

\bibitem{YangMills1954}
C.~N.~Yang and R.~L.~Mills,
\textit{Conservation of Isotopic Spin and Isotopic Gauge Invariance},
Phys. Rev. \textbf{96} (1954) 191--195.

\bibitem{KhabarovZinoviev2019}
M.~V.~Khabarov and Yu.~M.~Zinoviev,
\textit{Massive higher spin fields in the frame-like multispinor formalism},
Nucl. Phys. B \textbf{948} (2019) 114773,
arXiv:1906.03438 [hep-th].

\bibitem{Weinberg1995}
S.~Weinberg,
\textit{The Quantum Theory of Fields, Vol.~I: Foundations},
Cambridge University Press, Cambridge (1995).

\bibitem{Penrose1969}
R.~Penrose,
\textit{Solutions of the Zero-Rest-Mass Equations},
J. Math. Phys. \textbf{10} (1969) no.~1, 38--39,
doi:10.1063/1.1664756.

\bibitem{PeskinSchroeder1995}
M.~E.~Peskin and D.~V.~Schroeder,
\textit{An Introduction to Quantum Field Theory},
Addison--Wesley, Reading, MA (1995).

\bibitem{Wald1994}
R.~M.~Wald,
\textit{Quantum Field Theory in Curved Spacetime and Black Hole Thermodynamics},
University of Chicago Press, Chicago (1994).

\bibitem{FewsterHunt2013}
C.~J.~Fewster and D.~S.~Hunt,
\textit{Quantization of linearized gravity in cosmological vacuum spacetimes},
Rev. Math. Phys. \textbf{25} (2013) 1330003,
arXiv:1203.0261 [math-ph],
doi:10.1142/S0129055X13300033.

\bibitem{HackSchenkel2013}
T.-P.~Hack and A.~Schenkel,
\textit{Linear bosonic and fermionic quantum gauge theories on curved spacetimes},
Gen. Relativ. Gravit. \textbf{45} (2013) 877--910,
arXiv:1205.3484 [math-ph],
doi:10.1007/s10714-013-1508-y.
\end{thebibliography}
\end{document}